\documentclass[11pt,titlepage,a4paper]{article}
\usepackage{bozhomac}	
\usepackage{bozhlogo}	
\usepackage{cite}

\title{\bfseries	\vspace*{-2.218in}
\vspace*{-3ex}
{
\begin{flushright}
	  \textbf{\large LANL xxx E-print archive No. quant-ph/0105056}\\[2ex]
\end{flushright}
}
{\huge Fibre bundle formulation of \\[0.22ex] relativistic quantum mechanics}
\\[1.1ex]
{\LARGE  I. Time-dependent approach}
}

\vspace{1.7ex}

\author{
Bozhidar Z. Iliev
\thanks{Department Mathematical Modeling,
Institute for Nuclear Research and \mbox{Nuclear} Energy,
Bulgarian Academy of Sciences,
Boul. Tzarigradsko chauss\'ee~72, 1784 Sofia, Bulgaria}
\thanks{E-mail address: bozho@inrne.bas.bg}
\thanks{URL: http://theo.inrne.bas.bg/$\sim$bozho/}
}

\date{
 \vspace{2.27ex}\ShortTitle{Bundle relativistic quantum mechanics: I}
								\\[0.27ex]
 \vspace{3.27ex}
	\begin{tabular}{r@{$\colon\to~$}l}
 \vspace{0.09ex} Basic ideas	& November 1997, January 1998	\\[0.09ex]
 \vspace{0.09ex} Began/Ended	& February 16, 1998/March 27, 1998\\[0.09ex]
 \vspace{0.09ex} Initial typeset& April 1--14, 1998	\\[0.09ex]
 \vspace{0.09ex} Revised	& August 1999		\\[0.09ex]
  \vspace{0.09ex} Last update	& May 12, 2001		\\[1.09ex]
\vspace{0.27ex} Composing/Extracting part I
			& April 21/April 24, 1998	\\[0.27ex]
\vspace{0.27ex}  Updating part I
			& April, May 8, October 1998;
  				August 1999		\\[0.27ex]
  \vspace{0.27ex} Produced	& \fbox{\today}		\\[0.27ex]
	\end{tabular} \\[1.27ex]
	\begin{tabular}{r@{$\colon~$}l}
\vspace{0.27ex} LANL xxx archive server E-print No. & quant-ph/0105056
 						\\[0.27ex]
	\end{tabular} \\[-0.27ex]
 \vspace{4.27ex}{\Huge\BOZHO}	\\[4.27ex]
 \vspace{0.27ex}\Subject{Relativistic quantum mechanics,
			  Differential geometry}		\\[2.27ex]
	\begin{tabular}{r@{\hspace{0.512em}}|@{\hspace{0.512em}}l}
 \vspace{0.27ex}		
&
 \vspace{0.27ex}
	\end{tabular} \\[1.27ex]
 \vspace{0.27ex}	\\[0.27ex]
}

\newcommand{\fibre}{\mathcal{F}} 
\newcommand{\bundle}{(\bspace,\pr,\basesp)}	
	\newcommand{\bspace}{\mathnormal{F}}	
	\newcommand{\pr}{\pi}			
	\newcommand{\basesp}{\mathnormal{M}}	
\newcommand{\fibreover}[1]{{\bspace_{#1}}} 

\newcommand{\Hil}{\mathcal{F}}	
\newcommand{\HilB}{(\bHil,\proj,\base)}	
	\newcommand{\bHil}{\mathit{F}}	
	\newcommand{\proj}{\pi}		
	\newcommand{\base}{\mathit{M}}	

\newcommand{\Ham}{\mathcal{H}}	
\newcommand{\bHam}{\mathit{H}}	

\newcommand{\mbHam}{\Mat{\bHam}^\mathbf{m}} 

\newcommand{\dyn}[1]{\pmb{\mathbb{#1}}}	
	\newcommand{\ope}[1]{\mathcal{#1}}		 
	\newcommand{\mor}[1]{\mathit{#1}}		 
	\newcommand{\mmor}[1]{\Mat{\mathit{#1}}}	 

\begin{document}		

\renewcommand{\thefootnote}{\fnsymbol{footnote}} 
\maketitle				
\renewcommand{\thefootnote}{\arabic{footnote}}   

\tableofcontents		


\pagestyle{myheadings}
\markright{\itshape\bfseries Bozhidar Z. Iliev:~~
	\upshape\sffamily\bfseries Bundle relativistic quantum mechanics. I}


\begin{abstract}

	We propose a new fibre bundle formulation of the mathematical base of
relativistic quantum mechanics. At the present stage the bundle form of the
theory is equivalent to its conventional one, but it admits new types of
generalizations in different directions.

	In the present first part of our investigation we consider the
time\ndash dependent or Hamiltonian approach to bundle description of
relativistic quantum mechanics.
	In it the wavefunctions are replaced with (state) liftings of
paths or sections along paths of a suitably chosen vector bundle over
space\nobreakdash-time whose (standard) fibre is the space of the
wavefunctions. Now the quantum evolution is described as a linear
transportation (by means of the evolution transport along paths in the
space-time) of the state liftings/sections in the (total) bundle space. The
equations of these transportations turn to be the bundle versions of the
corresponding relativistic wave equations.

\end{abstract}

\section {Introduction}
\label{I.Introduction}

	This investigation is devoted to the fibre bundle (re)formulation of
the (mathematical) base of relativistic quantum mechanics. It can be regarded
as a direct continuation
of~\cite{bp-BQM-preliminary,
	bp-BQM-introduction+transport,bp-BQM-equations+observables,
	bp-BQM-pictures+integrals,bp-BQM-mixed_states+curvature,
	bp-BQM-interpretation+discussion,
	bp-BQM-full}
where a full self-consistent fibre bundle version of non-relativistic quantum
mechanics is elaborated. Many ideas of these works can be transferred to the
relativistic case but, as we shall see, they are not enough for the
above aim. For example, the right mathematical concept reflecting the
peculiarities of relativistic quantum evolution is the (linear) transport
along the identity map of the space-time, on the contrary to a (linear)
transport along paths (in space-time) in the non-relativistic case. A new
problem is the fact that some relativistic wave equations are linear partial
differential equations of second order; so for them the methods developed
for the treatment of Schr\"odinger equation can not be applied directly.

	In the bundle description the relativistic wavefunctions are replaced
by (state) liftings of paths or sections along paths of a suitable vector
bundle with the space-time as a base and whose (standard, typical) fibre is
the space where the corresponding conventional wavefunctions (or some
combinations of them and their partial derivatives) `live'. The operators
acting on this space are replaced with appropriate liftings of paths
or morphisms along paths and the evolution of the wavefunctions is now
described as a linear transport (along the identity map of the space-time or
along paths in it) of the state liftings/sections.

	The present first part of our work is devoted to the time-dependent
or Hamiltonian bundle description of relativistic wave equations. This
approach is a straightforward generalization of the bundle description of
nonrelativistic quantum mechanics to relativistic one.

	The lay-out of the paper is the following.

	A review of the bundle approach to non-relativistic quantum
mechanics is presented in Sect.~\ref{Sect2}. It contains certain basic ideas
and equations
of~\cite{bp-BQM-introduction+transport,bp-BQM-equations+observables,
	bp-BQM-pictures+integrals,bp-BQM-mixed_states+curvature,
	bp-BQM-interpretation+discussion,bp-BQM-full}
required as a starting point for the present work.

	In Sect.~\ref{Sect3} is developed a general scheme for fibre
bundle treatment of Schr\"odinger-type partial differential equations. The
method is also applicable to linear partial differential equations of higher
orders.  This is archived by transforming such an equation to a system of
first-order linear partial differential equations (with respect to a new
function) which, when written in a matrix form, is just the
Schr\"odinger-like presentation of the initial equation. Of course, this
procedure is not unique and its concrete realization depends on the physical
problem under exploration.

	In Sect.~\ref{Sect4} is given a bundle description of Dirac
equation.  Now the corresponding vector bundle over the space-time has
as a (standard) fibre the space of 4-spinors and may be called
4-spinor bundle. Since Dirac equation is of first order, it can be rewritten
in Schr\"odinger-type form. By this reason the formalism outlined in
Sect.~\ref{Sect2} is applied to it \emph{mutatis mutandis}. In particular,
here the state of a Dirac particle is described by a lifting of paths or
section along paths of the 4-spinor bundle and equivalently rewrite Dirac
equation as an equation for linear transportation of this lifting with
respect to the corresponding \emph{Dirac evolution transport} in this bundle.

	Sect.~\ref{Sect5} contains several procedures for transforming
Klein-Gordon equation to Schr\"odinger-like one. After such a presentation is
chosen, we can, analogously to Dirac case in Sect.~\ref{Sect4}, apply
to it the methods of Sect.~\ref{Sect2} and Sect.~\ref{Sect3}.

	Comments on fibre bundle description of other relativistic wave
equations are given in Sect.~\ref{Sect6}.

	Sect.~\ref{I.Conclusion} closes the paper with some remarks.

\section
[Bundle nonrelativistic quantum mechanics (review)]
{Bundle nonrelativistic quantum mechanics\\ (review)}
\label{Sect2}

	In the series of
papers~\cite{bp-BQM-introduction+transport,bp-BQM-equations+observables,
      bp-BQM-pictures+integrals,bp-BQM-mixed_states+curvature,
      bp-BQM-interpretation+discussion,bp-BQM-full}
we have reformulated nonrelativistic quantum mechanics in terms of fibre
bundles. The mathematical base for this was the Schr\"odinger equation
	\begin{equation}	\label{2.1}
\ih\frac{\ordinary\psi(t)}{\ordinary t} = \Ham(t) \psi(t),
	\end{equation}
where $\iu\in\mathbb{C}$ is the imaginary unit, $\hbar(=h/2\pi)$ is the Plank
constant divided by $2\pi$, $\psi$ is the system's state vector belonging to
an appropriate Hilbert space $\Hil$, and $\Ham$ is the system's Hamiltonian.
Here $t$ is the time considered as an independent variable (or parameter). If
$\psi$ is known for some initial moment $t_0$, the solution of~\ref{2.1} can
be written as
	\begin{equation}	\label{2.2}
\psi(t) = \ope{U}(t,t_0)\psi(t_0)
	\end{equation}
where $\ope{U}$ is the \emph{evolution operator} of the
system~\cite[chapter~IV, sect.~3.2]{Prugovecki-QMinHS} (for details
see~\cite{bp-BQM-introduction+transport}).

	In the bundle approach the system's Hilbert space $\Hil$ is replace
with a Hilbert bundle $\HilB$ with (total, fibre) bundle space $\bHil$,
projection $\proj$, base $\base$, isomorphic fibres
$\bHil_x:=\proj^{-1}(x)$, $x\in\base$, and (standard, typical) fibre
coinciding with $\Hil$. So, there exist isomorphisms
$l_x\colon\bHil_x\to\Hil$, $x\in\base$.
	In the present work the base $\base$ will be identified with the
Minkowski space-time $M^4$ of special relativity.%
\footnote{%
The bundle formulation of nonrelativistic quantum mechanics is insensitive to
the/this choice of $\base$~\cite{bp-BQM-interpretation+discussion}; in it is
more natural to identify $\base$ with the 3-dimensional Newtonian space
$\mathbb{E}^3$ of classical mechanics.%
}

	In the \emph{Hilbert bundle description} a state vector $\psi$ and
the Hamiltonian $\Ham$ are replaced respectively by a state lifting of
paths $\Psi\colon\gamma\to\Psi_\gamma$ and the \emph{bundle Hamiltonian}
(morphism along paths) $\bHam\colon\gamma\to\bHam_\gamma$, given
by~\cite{bp-BQM-introduction+transport,bp-BQM-equations+observables,
      bp-BQM-interpretation+discussion}:
	\begin{equation}	\label{2.3}
\Psi_\gamma\colon t\to \Psi_\gamma(t) = l_{\gamma(t)}^{-1}\bigl(\psi(t)\bigr),
\qquad
\bHam_\gamma\colon t\to
\bHam_\gamma(t) = l_{\gamma(t)}^{-1} \circ \Ham(t) \circ l_{\gamma(t)}
	\end{equation}
where $\gamma\colon J\to\base$, $J$ being an $\mathbb{R}$-interval, is the
world line (path) of some (point-like) observer, $t\in J$, and $\circ$
denotes composition of maps.

	The bundle analogue of the evolution operator $\ope{U}$ is the
\emph{evolution transport} $\mor{U}$ along paths, both connected by
	\begin{equation}	\label{2.4}
\mor{U}_\gamma(t,s)
=l_{\gamma(t)}^{-1}\circ \ope{U}(t,s) \circ l_{\gamma(s)}
\colon\bHil_{\gamma(s)}\to\bHil_{\gamma(t)},
\qquad s,t\in J,
	\end{equation}
which governs the evolution of state liftings via (cf.~\eref{2.2})
	\begin{equation}	\label{2.5}
\Psi_\gamma(t) = \mor{U}_\gamma(t,s)\Psi_\gamma(s), \qquad s,t\in J.
	\end{equation}

	The bundle version of~\eref{2.1}, the so-called
\emph{bundle Schr\"odinger equation},
is~\cite{bp-BQM-equations+observables,bp-BQM-full}
	\begin{equation}	\label{2.6}
\mor{D}\Psi = 0 .
	\end{equation}
Here $\mor{D}$ is the derivation
along paths corresponding to $\mor{U}$, viz (cf.~\cite{bp-normalF-LTP}
or~\cite[definition~4.1]{bp-LTP-general};
see also~\cite[definition~3.4]{bp-BQM-introduction+transport})
	\begin{equation*}
	D\colon\PLift^1(E,\pi,B) \to \PLift^0(E,\pi,B)
	\end{equation*}
where $\PLift^k\HilB$ is the set of $C^k$ liftings of paths from $\base$ to
$\bHil$, and its action on a lifting $\lambda\in\PLift^1\HilB$ with
$\lambda\colon\gamma\mapsto\lambda_\gamma$ is given via
	\begin{equation}	\label{2.7}
\mor{D}_{s}^{\gamma}\lambda :=
  \lim_{\varepsilon\to 0}
\left\{  \frac{1}{\varepsilon}
 \bigl[
\mor{U}_\gamma(s,s+\varepsilon)\lambda_\gamma(s+\varepsilon)
	- \lambda_\gamma(s)
 \bigr]
\right\}
	\end{equation}
where
$D_{s}^{\gamma}(\lambda):=((D\lambda)(\gamma))(s)=(D\lambda)_\gamma(s)$.

	If $\{e_a(\gamma(s))\}$, $s\in J$ is a basis in
$\fibreover{\gamma(s)}$, the explicit action of $\mor{D}$
is~\cite[proposition~4.2]{bp-LTP-general} (see also~\cite{bp-normalF-LTP})
	\begin{equation}	\label{2.8}
\mor{D}_{s}^{\gamma}\lambda =
\left(
\frac{\od\lambda^a_\gamma(s)}{\od s} +
\Gamma_{{\ }b}^{a}(s;\gamma)\lambda^b_\gamma(s)
\right)
e_a(\gamma(s)) .
	\end{equation}
Here the
\emph{coefficients}
$\Gamma_{{\ }a}^{b}(s;\gamma)$ of $\mor{U}$ are defined by
	\begin{equation}	\label{2.9}
\Gamma_{{\ }a}^{b}(s;\gamma) :=
\left.
\frac{\partial\left(\mor{U}_\gamma(s,t)\right)_{{\ }a}^{b}}{\partial t}
\right|_{t=s} =
- \left.
\frac{\partial\left(\mor{U}_\gamma(t,s)\right)_{{\ }a}^{b}}{\partial t}
\right|_{t=s}
	\end{equation}
where $\left(\mor{U}_\gamma(s,t)\right)_{{\ }a}^{b}$ are given via
\(
U(t,s)e_a(\gamma(s)) =:
	\sum_{b} \left(\mor{U}_\gamma(s,t)\right)_{{\ }a}^{b} e_b(\gamma(t))
\)
and are the local components of $U$ in $\{e_a\}$.

	There is a bijective correspondence between $\mor{D}$ and the
(bundle) Hamiltonian expressed by%
\footnote{%
We denote the matrix corresponding to some quantity, e.g.\ vector or
operator, in a given field of bases with the same (kernel) symbol but in
\textbf{boldface}; e.g.\
\(
\Mat{U}_\gamma(t,s)
:=
\bigl[ \left(\mor{U}_\gamma(s,t)\right)_{{\ }a}^{b} \ \bigr] .
\)%
}
	\begin{equation}	\label{2.10}
\Mat{\Gamma}_\gamma(t) :=
\bigl[ \Gamma_{{\ }a}^{b}(t;\gamma) \bigr] =
- \iih \mbHam_{\gamma}(t)
	\end{equation}
with
	\begin{equation*}
\mbHam_{\gamma}(t) =
\ih
\frac{\partial\mmor{U}_\gamma(t,t_0)}{\partial t}
\mmor{U}_\gamma^{-1}(t,t_0) =
\frac{\partial\mmor{U}_\gamma(t,t_0)}{\partial t}
\mmor{U}_\gamma^{}(t_0,t).
	\end{equation*}
being the \emph{matrix-bundle Hamiltonian} (for details
see~\cite{bp-BQM-equations+observables})%
\footnote{%
Note, the constant $\ih$ in~\eref{2.10} comes from the same constant
in~\eref{2.1}.%
}

	In the Hilbert space description of quantum mechanics to a dynamical
variable $\dyn{A}$ corresponds an observable $\ope{A}(t)$ which is a linear
Hermitian operator in $\Hil$. In the Hilbert bundle description to  $\dyn{A}$
corresponds a Hermitian lifting $\mor{A}$ of paths whose
restriction on $\bHil_{\gamma(t)}$ is
	\begin{equation}	\label{2.11}
\mor{A}_{\gamma}(t) =
l_{\gamma(t)}^{-1} \circ \ope{A}(t) \circ l_{\gamma(t)}
\colon  \bHil_{\gamma(t)}\to\bHil_{\gamma(t)}.
	\end{equation}

	The mean value of $\dyn{A}$ at a state characterized by a state
vector $\psi$ or, equivalently, by the corresponding to it
state lifting $\Psi_\gamma$ is
	\begin{equation}	\label{2.12}
\langle\ope{A}(t)\rangle_\psi^t
:= \frac{\langle\psi(t) | \ope{A}(t)\psi(t)\rangle}
	{\langle\psi(t) | \psi(t)\rangle}
= \left\langle \mor{A}_{\gamma}(t) \right\rangle_{\Psi_\gamma}^{t}
:=
\frac{
\langle \Psi_\gamma(t) | \mor{A}_\gamma(t) \Psi_\gamma(t) \rangle_{\gamma(t)}
}
{
\langle \Psi_\gamma(t) | \Psi_\gamma(t) \rangle_{\gamma(t)}
}.
	\end{equation}
Here
	\begin{equation}	\label{2.13}
\langle \cdot | \cdot \rangle_x =
\langle l_x \cdot | l_x \cdot \rangle, \qquad x\in\base
	\end{equation}
is the fibre Hermitian scalar product in $\HilB$ induced by  the Hermitian
scalar product
$\langle \cdot | \cdot \rangle \colon \Hil\times\Hil\to\mathbb{C}$
in $\Hil$.

	A summary of the above and other details concerning the Hilbert
space and Hilbert bundle description of (nonrelativistic) quantum mechanics
can be found in~\cite{bp-BQM-interpretation+discussion}.

\section{General case}
\label{Sect3}

	Consider now the pure mathematical aspects of the scheme described in
Sect.~\ref{Sect2}.
	On one hand, it is essential to be notice that the Schr\"odinger
equation~\eref{2.1} is a first order (with respect to the time%
\footnote{%
The dependence of $\psi$ and $\Ham$ on the spatial coordinates (and momentum
operators) is inessential for the present part of our investigation and,
respectively, is not written explicitly.%
}%
)
linear partial differential equation solved with respect to the time
derivative. On the other hand, the considerations of Sect.~\ref{Sect2} are
true for Hilbert spaces and bundles whose dimensionality is generically
infinity, but, as one can easily verify, they hold also for spaces and
bundles with finite dimension.

	These observations, as we shall prove below, are enough to transfer
the bundle nonrelativistic formalism to the relativistic region. We call the
result of this procedure
\emph{time\ndash dependent or Hamiltonian approach} as in
it the time plays a privileged r\^ole and the relativistic covariance is
implicit.

	Taking into account the above, we can make the following conclusion.
Given a linear (vector) space $\fibre$ of $C^1$ functions
$\psi\colon J \to \mathbb{C}$ with $J$ being an
$\mathbb{R}$\ndash interval. (We do
not make any assumptions on the dimensionality of $\fibre$; it can be finite
as well as countable or uncountable infinity.) Let
$\Ham(t)\colon\fibre\to\fibre$, $t\in J$ be (possibly depending on $t$)
linear operator. Consider the equation%
\footnote{%
We introduce the multiplier $\ih\not=0$ from purely physical reasons and to be
able to apply the results already obtained directly, without any changes.%
}
	\begin{equation}	\label{3.1}
\ih\frac{\pd \psi(t)}{\pd t} = \Ham(t) \psi(t).
	\end{equation}
For it are valid all of the results of Sect.~\ref{Sect2}, viz., for instance,
there can be introduced the bundle $\bundle$, the evolution operator
$\ope{U}$, etc.; the relations between them being the same as in
Sect.~\ref{Sect2}. If the vector space $\fibre$ is endowed with a scalar
(inner) product
$\langle \cdot | \cdot \rangle \colon \fibre\times\fibre\to\mathbb{C}$,
then~\eref{2.13} induces analogous product in the bundle (\ie in the
bundle's fibres).
Consequently, imposing condition~\eref{2.12}, we get~\eref{2.11}. Further,
step by step, one can derive all of the results
of~\cite{bp-BQM-introduction+transport,bp-BQM-equations+observables,
	bp-BQM-pictures+integrals,bp-BQM-mixed_states+curvature,
	bp-BQM-interpretation+discussion,bp-BQM-full}.

	Further we shall need a slight, but very important generalization of
the above. Let now $\fibre$ be a vector space of vector-valued $C^1$ functions
$\psi\colon J \to V$ with $V$ being a complex vector space and
$\Ham(t)\colon\fibre\to\fibre$; the case just considered corresponds to
$V=\mathbb{C}$, i.e to $\dim_\mathbb{C}V=1$. It is not difficult to verify
that all of the above-said, corresponding to $\dim_\mathbb{C}V=1$, is also
\emph{mutatis mutandis} valid for $\dim_\mathbb{C}V\ge1$. Consequently, the
fibre bundle reformulation of the solution of~\eref{3.1}, the operators, and
scalar product(s) in $\fibre$ can be carried out in the general case when
$\psi\colon J\to V$ with $\dim_\mathbb{C}V\ge1$.

	Suppose now $\mathit{K}^{m}$ is the vector space of $C^m$,
$m\in\mathbb{N}\cup\{\infty\}\cup\{\omega\}$, vector-valued functions
$\varphi\colon J\to W$ with $W$ being a vector space. Let
$\varphi\in\mathit{K}^{m}$ satisfies the equation
	\begin{equation}	\label{3.2}
f\left(
t,\varphi,\frac{\pd\varphi}{\pd t},\ldots,\frac{\pd^n\varphi}{\pd t^n}
\right)
= 0
	\end{equation}
where
 \(
f\colon J\times
\mathit{K}^{m}\times\mathit{K}^{m-1}\times\cdots\times\mathit{K}^{m-n}
\to \mathit{K}^{m-n},
 \)
 $n\in\mathbb{N},\ n\le m$
is a map (multi)linear in $\varphi$ and its derivatives. We suppose~\eref{3.2}
to be solvable with respect to the highest derivative of $\varphi$,
i.e.~\eref{3.2} to be equivalent to%
\footnote{%
Requiring the equivalence of~\eref{3.2} and~\eref{3.3}, we exclude the
existence of sets on which the highest derivative of $\varphi$ entering
in~\eref{3.2} may be of order $k<n$. If we admit the existence of such sets,
then on them we have to replace $n$ in~\eref{3.3} with $k$. (Note $k$ may be
different for different such sets.) The below-described procedure can be
modified to include this more general situation but, since we do not want to
fill the presentation with complicated mathematical details, we are not going
to do this here.%
}
	\begin{equation}	\label{3.3}
\frac{\pd^n\varphi}{\pd t^n}
=
G\left(
t,\varphi,\frac{\pd\varphi}{\pd t},\ldots,\frac{\pd^{n-1}\varphi}{\pd t^{n-1}}
\right)
       	\end{equation}
for some map
\(
G\colon J\times
\mathit{K}^{m}\times\mathit{K}^{m-1}\times\cdots\times\mathit{K}^{m-n+1}
\to \mathit{K}^{m-n},
\)
linear in $\varphi$ and its derivatives.

	The already developed fibre bundle formalism for the
(Schr\"odinger-type) equation~\eref{3.1} can be transferred to~\eref{3.3}.
This can be done in a number of different ways. Below we shall realize the
most natural way, but one has to keep in mind that for a concrete equation
another method may turn to be more useful, especially from the view-point of
possible physical applications (see below Sect.~\ref{Sect5}).

	Defining
\(
\fibre :=
\mathit{K}^{m}\times\mathit{K}^{m-1}\times\cdots\times\mathit{K}^{m-n+1}
\)
and putting
	\begin{equation}	\label{3.3-1}
\psi :=
\left(
\varphi,\frac{\pd\varphi}{\pd t},\ldots,\frac{\pd^{n-1}\varphi}{\pd t^{n-1}}
\right)^\top
	\end{equation}
with $\top$ being the transposition sign, we can transform~\eref{3.3} in the
form~\eref{3.1} with `Hamiltonian'
	\begin{equation}	\label{3.4}
	\begin{split}
       & \Ham(t) = \ih \times
\\
&\mspace{-2.1mu}
\begin{pmatrix}
0 & \id_{\mathit{K}^{m-1}} & 0                      & \ldots &  0 \\
0 & 0                      & \id_{\mathit{K}^{m-2}} & \ldots &  0 \\
\hdotsfor[2.11]{5} \\
0 & 0                      & 0                      & \ldots &
						\id_{\mathit{K}^{m-n+1}} \\
f_0(t)\id_{\mathit{K}^{m}} &
	f_1(t)\id_{\mathit{K}^{m-1}} &
		f_2(t)\id_{\mathit{K}^{m-2}} &
					\ldots &
					f_{n-1}(t)\id_{\mathit{K}^{m-n+1}}
\end{pmatrix}
	\ \colon\fibre\to\fibre
	\end{split}
	\end{equation}
which is a linear matrix operator, \ie a matrix of
linear operators.%
\footnote{%
Operators of this kind will be considered in the next part of the present
investigation.%
}
Here $\id_X$ is the identity map of a set $X$ and
$f_i\colon J\to\mathbb{C}$, $i=0,\ldots,n-1$ define the (multi)linear map
$G\colon J\times\fibre\to\mathit{K}^{m-n}$ by
	\begin{equation}	\label{3.5}
G\left(
t,\varphi,\frac{\pd\varphi}{\pd t},\ldots,\frac{\pd^{n-1}\varphi}{\pd t^{n-1}}
\right)
=
f_0(t)\varphi +
\sum_{i=1}^{n-1}f_i(t)\frac{\pd^i\varphi}{\pd t^i}.
	\end{equation}

	In this way we have proved that~\eref{3.2} can equivalently be
rewritten as a Schr\"odinger-type equation~\eref{3.1} with `Hamiltonian'
given by~\eref{3.4}. Such a transformation is not unique. For example, one
can choose the components of $\psi$ to be any $n$ linearly independent linear
combinations of
$\varphi$, $\pd\varphi/\pd t$, \ldots,  $\pd^{n-1}\varphi/\pd t^{n-1}$; this
will result only in another form of the matrix~\eref{3.4}.
In fact, if $A(t)$ is a nondegenerate matrix\ndash valued function, the
change
	\begin{equation}	\label{3.7}
\psi(t)\mapsto\widetilde{\psi}(t) = A(t)\psi(t)
	\end{equation}
leads to~\eref{3.1} with Hamiltonian
	\begin{equation}	\label{3.8}
\widetilde{\Ham}(t) = A(t)\Ham A^{-1}(t) +\frac{\pd A(t)}{\pd t}A^{-1}(t) .
	\end{equation}

	Now to the equation~\eref{3.1}, with $\Ham$ given via~\eref{3.4}, we
can apply the already described procedure for reformulation in terms of
bundles.

\section {Dirac equation}
\label{Sect4}

	The relativistic quantum mechanics of spin $\frac{1}{2}$ particle is
described by the Dirac
equation (see~\cite[chapter~2]{Itzykson&Zuber},
\cite[chapters~1\nobreakdash--5]{Bjorken&Drell-1},
and~\cite[part~V, chapter~XX]{Messiah-2}).
The wave function $\psi=(\psi_1,\psi_2,\psi_3,\psi_4)^\top$ of such a
particle is a 4\ndash dimensional (4\ndash component) spinor satisfying this
equation whose Schr\"odin\-ger\ndash type form%
\footnote{%
For the purposes of this section we do not need the widely know relativistic
invariant form of Dirac
equation~\cite{Itzykson&Zuber, Bjorken&Drell-1, Messiah-2}.%
}
is (see, e.g.~\cite[chapter~XX, \S~6, equation~(36)]{Messiah-2}
or~\cite[chapter~1, \S~3, equation~(1.14)]{Bjorken&Drell-1})
	\begin{equation}	\label{4.1}
\ih\frac{\pd\psi}{\pd t} = \lindexrm[\Ham]{}{D} \psi
	\end{equation}
where $\lindexrm[\Ham]{}{D}$ is a Hermitian operator, called
\emph{Dirac Hamiltonian}, in the space $\fibre$ of state vectors (spinors).
For a spin $\frac{1}{2}$ particle with (proper) mass $m$ and electric charge
$e$ the explicit form of $\lindexrm[\Ham]{}{D}$ in an (external)
electromagnetic field with 4\ndash potential $(\varphi,\Vect{\mathit{A}})$
is~\cite[chapter~XX, \S~6, equation~(44)]{Messiah-2}
	\begin{equation}	\label{4.2}
\lindexrm[\Ham]{}{D} =
e\varphi\openone_4 +
c\Vect{\alpha} \cdot
	(\Vect{p} - \frac{e}{c} \Vect{\mathit{A}} ) + mc^2\beta,
	\end{equation}
where $\openone_4$ is the $4\times4$ unit matrix,
$c$ is the velocity of light in vacuum,
$\Vect{\alpha}:=(\alpha^1,\alpha^2,\alpha^3)$ is a matrix vector of
$4\times4$ matrices, $\Vect{p}=-\ih\boldsymbol{\nabla}$ is the
(3\ndash dimensional) momentum operator ($\nabla_i:=\pd/\pd x^i,\ i=1,2,3$),
and $\beta$ is a $4\times4$ matrix. The explicit (and general) forms of
 $\alpha^1,\alpha^2,\alpha^3$, and $\beta$ can be found, e.g.
in~\cite[chapter~2, sect.~2.1.2, equation~(2.10)]{Itzykson&Zuber}.

	Since~\eref{4.1} is a first-order equation, we can introduce the
\emph{Dirac evolution operator} $\lindexrm[\ope{U}]{}{D}$ via
 $\psi(t)=\lindexrm[\ope{U}]{}{D}(t,t_0)\psi(t_0)$ (cf.~\eref{2.2}).
Generally it is a $4\times4$ integral\ndash matrix operator uniquely
defined by the initial\ndash value problem
	\begin{equation}	\label{4.3}
\ih\frac{\partial}{\partial t} \lindexrm[\ope{U}]{}{D}(t,t_0) =
\lindexrm[\Ham(t)]{}{D} \circ \lindexrm[\ope{U}]{}{D}(t,t_0),
\qquad
\lindexrm[\ope{U}]{}{D}(t_0,t_0) = \id_\fibre
	\end{equation}
with $\fibre$ being the space of 4-spinors.
The explicit form of $\lindexrm[\ope{U}]{}{D}$ is derived, e.g.\
in~\cite[chapter~2, sect.~2.5.1]{Itzykson&Zuber}.

	Now the bundle formalism developed
in~\cite{bp-BQM-introduction+transport,bp-BQM-equations+observables,
	bp-BQM-pictures+integrals,bp-BQM-mixed_states+curvature,
	bp-BQM-interpretation+discussion,bp-BQM-full}
can be applied to a description of Dirac particles practically without
changes. For instance, the spinor lifting of paths have to be introduced
via~\eref{2.3} and are connected by~\eref{2.5} in which $\mor{U}$ has to be
replaced by the \emph{Dirac evolution transport}
$\lindexrm[\mspace{-2.2mu}\mor{U}]{}{D}$ given
by
	\begin{equation*}
\lindexrm[\mspace{-2.2mu}\mor{U}]{}{D}_\gamma(t,s)
=
l_{\gamma(t)}^{-1}\circ \lindexrm[\ope{U}]{}{D}(t,s) \circ l_{\gamma(s)},
\qquad s,t\in J
	\end{equation*}
(cf.~\eref{2.4}). The \emph{bundle Dirac equation} is
	\begin{equation}	\label{4.4}
\lindexrm[\mspace{-2.2mu}\mor{D}]{}{D}_{t}^{\gamma} \Psi_\gamma = 0
	\end{equation}
with $\lindexrm[\mspace{-2.2mu}\mor{D}]{}{D}$ being the assigned to
$\lindexrm[\mspace{-2.2mu}\mor{U}]{}{D}$ by~\eref{2.7} derivation along
paths. Again, the matrix of the coefficients of the Dirac evolution transport
is connected with the matrix-bundle Dirac Hamiltonian via~\eref{2.10}, etc.

\section {Klein-Gordon equation}
\label{Sect5}

	The wavefunction $\phi\in\mathit{K}^m$ of spinless special\ndash
relativistic particle is a scalar function of class $C^m,\ m\geq2$, over the
spacetime and satisfies the Klein-Gordon
equation~\cite[chapte~9]{Bjorken&Drell-1}.  For a particle of mass $m$ and
electric charge $e$ in an external electromagnetic field with 4\ndash
potential $(\varphi,\Vect{\mathit{A}})$ it
reads~\cite[chapter~XX, \S~5, equation~(30)]{Messiah-2}
	\begin{equation}	\label{5.1}
\Bigl[
\Bigl(\ih\frac{\pd}{\pd t} -e\varphi\Bigr)^2
-
c^2 \left( \Vect{p} - \frac{e}{c}\Vect{\mathit{A}} \right)^2
\Bigr]
\phi
=m^2c^4\phi.
	\end{equation}

	This is a second\ndash order linear partial differential equation of
type~\eref{3.2} with respect to $\phi$. Solving it with respect to
$\pd^2\phi/\pd t^2$, we can transform it to equation of type~\eref{3.3}:
	\begin{equation}	\label{5.2}
	\begin{split}
&	\frac{\pd^2\phi}{\pd t^2}
=
f_0\phi + \frac{2e}{\ih}\varphi\frac{\pd\phi}{\pd t} \\
&	f_0 :=
\left[
-\frac{c^2}{\hbar^2}
\left( \Vect{p} - \frac{e}{c}\Vect{\mathit{A}} \right)^2
-
\frac{m^2c^4}{\hbar^2}
+
\frac{e^2}{\hbar^2}\varphi^2
+
\frac{2e}{\ih}\frac{\pd\varphi}{\pd t}
\right]
\id_{\mathit{K}^m}.
	\end{split}
	\end{equation}

	As pointed above, there are (infinitely many) different ways to put
this equation into Schr\"odinger\nobreakdash-type form. Below we realize
three of them, each having applications for different purposes.

	The `canonical' way to do this is to define
 $\psi := (\phi,\pd\phi/\pd t)^\top$ and
 \(
\mathit{K}^m :=
	\{ \phi\colon J\to\mathbb{C},\quad \phi\text{ is of class } C^m \},\
	m\geq2.
 \)
Then, comparing~\eref{5.2} with~\eref{3.5}, we see that~\eref{5.2}, and
hence~\eref{5.1}, is equivalent to~\eref{3.1} with
$\Ham=\lindexrm[\Ham]{\mspace{32mu}c}{K-G}$, where the `canonical'
Klein\ndash Gordon Hamiltonian is
	\begin{equation}	\label{5.3}
\lindexrm[\Ham]{\mspace{32mu}c}{K-G}
:=
\ih
	\begin{pmatrix}
0   & \id_{\mathit{K}^{m}}			 \\
f_0 & \frac{2e}{\ih}\varphi\id_{\mathit{K}^{m-1}}
	\end{pmatrix}
~~~.
	\end{equation}
Note, for a free particle, $(\varphi,\Vect{\mathit{A}})=0$, this is the
anti\ndash diagonal matrix operator
\(
\left(\begin{smallmatrix}
0                            & 1  \\
c^2\nabla^2 - m^2c^4/\hbar^2 & 0
\end{smallmatrix}\right)
\id_{\mathit{K}^{m}}
\).

	Another possibility is to put
\(
\psi=\bigl( \phi + \frac{\ih}{mc^2}\frac{\pd\phi}{\pd t},\phi -
	\frac{\ih}{mc^2}\frac{\pd\phi}{\pd t} \bigr)^\top.
\)
This choice is good for investigation of the non-relativistic limit, when
$\ih\frac{\pd\phi}{\pd t}\approx mc^2\phi$~\cite[chapterr~XX,
\S~5]{Messiah-2}, so that in it $\psi\approx(2\phi,0)^\top$.

	Now, as a simple verification proves, the Schr\"odinger\ndash type
form~\eref{3.1} of~\eref{5.2} is realized for the Hamiltonian
	\begin{multline*}
\lindexrm[\Ham]{\mspace{13.0mu}n.r.}{K-G} := \frac{1}{2} \times \\
	\begin{pmatrix}
(mc^2 + 2e\varphi)\id_{\mathit{K}^{m-1}} - \hbar^2f_0/mc^2 &
	(- mc^2 - 2e\varphi)\id_{\mathit{K}^{m-1}} - \hbar^2f_0/mc^2 \\
(mc^2 - 2e\varphi)\id_{\mathit{K}^{m-1}} + \hbar^2f_0/mc^2 &
	(- mc^2 + 2e\varphi)\id_{\mathit{K}^{m-1}} + \hbar^2f_0/mc^2
	\end{pmatrix}~.
	\end{multline*}

	If the electromagnetic field vanishes,
 $(\varphi,\Vect{\mathit{A}})=0$, then
 $f_0=(c^2\nabla^2-m^2c^4/\hbar^2)\id_{\mathit{K}^{m}}$ and
	\begin{multline*}
\lindexrm[\Ham]{\mspace{13mu}n.r.}{K-G} = \frac{1}{2}
\left[
mc^2
	\begin{pmatrix}
1 &  0 \\
0 & -1
	\end{pmatrix}
- \frac{\hbar^2}{2m}\boldsymbol{\nabla}^2
	\begin{pmatrix}
1  &  1 \\
-1 & -1
	\end{pmatrix}
\right] \id_{\mathit{K}^{m-1}}
					\\
=
\left[
\Bigl( mc^2 + \frac{\Vect{p}^2}{2m} \Bigr) \mspace{3mu}
	\begin{pmatrix}
1  &  0 \\
0  & -1
	\end{pmatrix}
+ \frac{\Vect{p}^2}{2m}
	\begin{pmatrix}
0  &  1 \\
-1 & 0
	\end{pmatrix}
\right] \id_{\mathit{K}^{m-1}}
	\end{multline*}

	The third possibility mentioned corresponds to the choice of $\psi$
as a $5\times1$ matrix:
\[
\psi =
\bigl(
mc^2\phi, \frac{\pd\phi}{\pd t}, \frac{\pd\phi}{\pd x^1},
\frac{\pd\phi}{\pd x^2}, \frac{\pd\phi}{\pd x^3}
\bigr)^\top
=
\bigl( mc^2\phi, \frac{\pd\phi}{\pd t}, \boldsymbol{\nabla}\phi \bigr)^\top.
\]
The corresponding Hamiltonian is $5\times5$ matrix which in the absence of
electromagnetic field is~\cite[chapter~2, sect.~2.1.1]{Itzykson&Zuber}
\(
\lindexrm[\Ham]{\mspace{30.2mu}5}{K-G}
= mc^2\beta+\Vect{\alpha}\cdot\boldsymbol{\nabla}
\)
(cf.\ Dirac Hamiltonian~\eref{4.2} for $(\varphi,\Vect{\mathit{A}})=0$)
where $\beta$ is $5\times5$ matrix and $\Vect{\alpha}$ is
a 3\ndash vector of  $5\times5$ matrices. The full realization of this
procedure and the explicit form of the corresponding $5\times5$ matrices is
given in~\cite[\S~4, sect.~4.4]{Bogolyubov&Shirkov}.

	Now choosing some representation of Klein\ndash Gordon equation as
first\ndash order (Schr\"odinger\nobreakdash-type) equation, we can in an
evident way transfer the bundle formalism to the description of spinless
particles.

\section {Other relativistic wave equations}
\label{Sect6}

	As we saw in sections~\ref{Sect4} and~\ref{Sect5}, the only
problem for a bundle reformulation of a wave equation is to rewrite it as a
first\nobreakdash-order differential equation and to find the corresponding
Hamiltonian.  Since all relativistic wave equations are of the form of
equation~\eref{3.2}~\cite{Bogolyubov&Shirkov,Messiah-2,Bjorken&Drell-1}, this
procedure can successfully be performed for all of them.

	For instance, the wavefunction
$\psi=(\psi_0,\psi_1,\psi_2,\psi_3)^\top$ of particles with spin~1 is a
4\ndash vector satisfying Klein-Gordon equation%
\footnote{%
Here we do not concern the additional conditions like the Lorentz one. They
lead to a modification of the equations defining the evolution transport, but
this does not change the main ideas. E.g.\ the most general equation for
vectorial mesons is the Proca
equation~\cite[chapter~3, equation~(3.132)]{Itzykson&Zuber}
\(
\bigl[
\bigl( \partial^\varkappa\partial_\varkappa + \frac{m^2c^4}{\hbar^2} \bigr)
\eta_{\mu\nu}
-
\partial_\mu\partial_\nu
\bigr]
\phi^\nu
= 0
\)
on which one usually imposes the additional condition $\pd_\nu\phi^\nu=0$.
(For $m\not=0$ the last condition is a corollary of the Proca equation; to
prove this, simply apply $\pd^\mu$ to Proca equation.)%
}%
~\cite[capter~I, \S~4]{Bogolyubov&Shirkov}. Hence for each component
$\psi_i$, $i=0,1,2,3$ we can construct the corresponding Hamiltonian
$\lindexrm[\Ham]{}{K-G}_i$ using, e.g., one of the methods described in
Sect.~\ref{Sect5}.
Then equation~\eref{3.1} holds for a Hamiltonian of the form of a
$4\times4$ diagonal block matrix operator
\(
\lindexrm[\Ham]{}{K-G} =
	\diag\bigl( \lindexrm[\Ham]{}{K-G}_0, \lindexrm[\Ham]{}{K-G}_1,
	\lindexrm[\Ham]{}{K-G}_2, \lindexrm[\Ham]{}{K-G}_3 \bigr)
\)
and the corresponding new wavefunction which is now a $8\times1$ matrix.

	The just said is, of course, valid with respect to electromagnetic
field, the 4\ndash potential playing the r\^ole of
wavefunction~\cite[chapter~I, \S~5]{Bogolyubov&Shirkov}. It is interesting to
be noted that even at a level of classical electrodynamics the Maxwell
equations admit Schr\"odinger\nobreakdash-like form. There are different
fashions to do this.
For a free field, one of them is to put
\(
\psi = (\Vect{E},\Vect{H},\Vect{E},\Vect{H})^\top
\)
where $\Vect{E}$ and $\Vect{H}$ are respectively the electric and
magnetic field strengths, and to define the Hamiltonian, e.g., by
\(
\lindexrm[\Ham]{}{E-M} = \ih
	\left(	\begin{smallmatrix}
0	& c\rot	& 0	& 0	\\
-c\rot	& 0	& 0	& 0	\\
0	& 0	& 0	& \diver\\
0	& 0	&-\diver& 0
	\end{smallmatrix}	\right)~.
\)

	Almost the same as spin~1 particles is the case of particles with
spin~2. Their wavefunction is symmetric tensor field of second rank whose
components satisfy also the Klein\ndash Gordon equation~\cite[p.~24]{Nelipa}.

	The wavefunction of particles with spin $3/2$ is a
(4\nobreakdash-)spin\ndash vector whose `vector' components satisfy the Dirac
equation (and some additional conditions)~\cite[pp.~35\Ndash 36]{Nelipa}.
Therefore for them can be applied \emph{mutatis mutandis} the presented in
Sect.~\ref{Sect4} bundle formalism.

\section {Conclusion}
\label{I.Conclusion}

	In this paper we have developed the time-dependent (Hamiltonian)
approach for fibre bundle formulation of relativistic quantum mechanics. As
we saw, it is a straightforward generalization of the methods worked out for
bundle treatment of non\ndash relativistic quantum mechanics. The
generic scheme is to transform a relativistic wave equation into a
Schr\"odinger\ndash like form with corresponding Hamiltonian and then to
apply \emph{mutatis mutandis} the results already established for the
Schr\"odinger equation. A new moment in the relativistic region is that some
of the wave equations are (system(s) of) partial differential equations of
order no less than two. The transformation of such (a system of) equations to
a first order Schr\"odinger\ndash like equation (or system of equations) is,
of course, non\ndash unique. The choice of such representation is more a
physical than a mathematical subject and depends on the concrete problem
posed. It is clear that different Schr\"odinger\ndash like representations
lead to similar (equivalent) but different bundle description of one and the
same initial equation.

	A `bad' feature of the Hamiltonian approach to bundle description of
relativistic quantum mechanics, presented in this paper, is its explicit
time\ndash dependence. A consequence of this fact is the implicit covariance
of the bundle description obtained in this way. Evidently, such a situation
is not satisfactory from the view\nobreakdash-point of relativistic character
of the theory it represents. This naturally leads us to the idea of
explicit\ndash covariant bundle description of relativistic quantum
mechanics. It turns out that for the solution of this problem are not enough
the methods developed for the Schr\"odinger equation. The physical reason for
this is that in the relativistic wave equations is intrinsically incorporated
the fact of absence of world lines (trajectories) in a classical sense of the
quantum objects they describe. These problems will be investigated in the
second part of this work where the covariant bundle description of
relativistic quantum mechanics will be developed.


We shall end with some comments on the material of
  Sect.~\ref{Sect3}.
We saw there that the fibre bundle formalism developed for the solutions of
Schr\"odinger equation can successfully be applied for the solutions of
(systems of) linear ordinary differential equations. For this purpose the
system of equations, if they are of order greater than one, has to be
transformed into a system of first-order equations. It can always be written
in a Schr\"odinger\ndash{like} form~\eref{3.1} to which the developed
in~\cite{bp-BQM-introduction+transport,bp-BQM-equations+observables,
	bp-BQM-pictures+integrals,bp-BQM-mixed_states+curvature,
	bp-BQM-interpretation+discussion,bp-BQM-full}
bundle approach can be applied \textit{mutatis mutandis}.

	Therefore, in particular, to a system of linear ordinary (with respect
to `time') differential equations corresponds a suitable linear transport
along paths in an appropriately chosen fibre bundle. As most of the
fundamental equations of physics are expressed by such systems of equations,
they admit fibre bundle (re)formulation analogous to the one of
Schr\"odinger equation.

	An interesting consequence of this discussion is worth mentioning.
Suppose a system of the above-described type is an Ouler-Lagrange (system of)
equation(s) for some Lagrangian. Applying the outlined `bundle' procedure,
we see that to this Lagrangian corresponds some (evolution) linear transport
along paths in a suitable fibre bundle. The bundle and the transport are
practically (up to isomorphisms) unique if the Ouler-Lagrange equations are of
first order with respect to time. Otherwise there are different (but
equivalent) such objects corresponding to the given Lagrangian. Hence, some
Lagrangians admit description in terms of linear transports along paths. In
more details the correspondence between Lagrangians (or Hamiltonians) and
linear transports along paths will be explored elsewhere.

\addcontentsline{toc}{section}{References}
\bibliography{bozhopub,bozhoref}
\bibliographystyle{unsrt}

\addcontentsline{toc}{subsubsection}{This article ends at page}

\end{document}